\documentclass[conference]{IEEEtran}
\IEEEoverridecommandlockouts
\usepackage{cite}
\usepackage{amsmath,amssymb,amsfonts}
\usepackage{algorithmic}
\usepackage{textcomp}
\usepackage{xcolor}
\usepackage[utf8]{inputenc}
\usepackage{breqn}
\usepackage{subimages}
\usepackage{balance}

\def\BibTeX{{\rm B\kern-.05em{\sc i\kern-.025em b}\kern-.08em
    T\kern-.1667em\lower.7ex\hbox{E}\kern-.125emX}}
\begin{document}

\title{Non-local Operational Anisotropic Diffusion Filter
}

\author{\IEEEauthorblockN{Fábio A. M. Cappabianco}
\IEEEauthorblockA{\textit{Group for Inovation Based on Images and Signals (GIBIS)} \\
\textit{Department of Science and Technology} \\
\textit{Federal University of São Paulo}\\
São José dos Campos/SP, Brazil \\
fcappabianco@gmail.com}
\and
\IEEEauthorblockN{Petrus P. C. E. da Silva}
\IEEEauthorblockA{\textit{Department of Medical Phisics} \\
\textit{A.C. Camargo Cancer Center}\\
São Paulo/SP, Brazil \\
petruspaulo@gmail.com}
}

\maketitle

\begin{abstract}
High-frequency noise is present in several modalities of medical images. It originates from the acquisition process and may be related to the scanner configurations, the scanned body, or to other external factors. This way, prospective filters are an important tool to improve the image quality. In this paper, we propose a non-local weighted operational anisotropic diffusion filter and evaluate its effect on magnetic resonance images and on kV/CBCT radiotherapy images. We also provide a detailed analysis of non-local parameter settings. Results show that the new filter enhances previous local implementations and has potential application in radiotherapy treatments.
\end{abstract}


\section{Introduction}

Medical imaging is subject to several kinds of noise and artifacts coming from the acquisition and storage procedure~\cite{Beutel:2000}. The nature and appearance of the noise depend on distinct factors such as the imaging modality, the scanned body part, scanner parameters and positioning, among others~\cite{Mohan:2014,Diwakar:2018}.

Despite recent improvements to the acquisition process, high-frequency filters are essential to improve medical image quality. Nevertheless, their application is not a trivial task, since some of the original signals may be removed and new artifacts might be inserted to the image while removing the noise~\cite{Maggioni:2013, Rafsanjani:2017}.

There are three generations of high-frequency noise filters: the isotropic which were proposed since the beginning of digital medical imaging~\cite{Lindenbaum:1994,Petrou:2010}; the anisotropic -- which became popular in the 90s~\cite{Alvarez:1992} -- are still in use for its robustness to generating artifacts, and is present in some popular medical imaging tools such as FLS~\cite{Smith:1997} and 3D Slicer~\cite{Fedorov:2012}; and the non-local anisotropic filters which started to be developed around 2005 with the Non-Local Means (NLM) algorithm~\cite{Buades:2005}. Even though the third generation produces better results in terms of several metrics, in the context of medical imaging, one should be even more careful since they tend to insert more artifacts to the filtered image~\cite{Maggioni:2013}.

In this paper we focus on the implementation and testing of an operational third generation filter which is an extension of the anisotropic diffusion filter (ADF)~\cite{Perona:1990} for reducing addictive noise present in MRI and CT images. The proposed implementation is operational with respect to local and non-local conservativeness parameter. We also conduct an analysis of the non-local parameters in the models proposed in~\cite{Palma:2014}. Our experiments include synthetic MRI images to estimate parameter values and also a qualitative evaluation of kV/CBCT radiotherapy images by specialists.

The reminder of this paper is organized as follows: Section~\ref{sec:rel.works} presents a review of related works in high-frequency image segmentation; Section~\ref{sec:anisotropic} contains an explanation of ADF and its extensions; Section~\ref{sec:method} proposes our new filtering method; Section~\ref{sec:experiments} shows our quantitative and qualitative experiments; and Section~\ref{sec:conclusion} states our conclusions.

\section{Related Works}
\label{sec:rel.works}

The first generation of high-frequency noise filters are isotropic, that is, the same operation is applied to each image pixel despite its contents or position. Some of the traditional isotropic filters are the arithmetic and geometrical means, median, and alpha-trimmed mean filter~\cite{Petrou:2010}.

The second generation of high-frequency noise filters is applied over other transform domains and/or with local anisotropic functions. These filters apply distinct operations according to pixel features such as intensity frequency or local patterns. Wavelet transform and cosine transform based~\cite{Anand:2010, Manjon:2012}, bilateral and trilateral~\cite{Wong:2004, Xie:2008}, and the diffusion~\cite{Gerig:1992, Black:1998} are among the most popular filters. While first generation filters were simpleton implementations, the second generation filters bear several non-trivial parameters which yielded lots of tuning and evaluation papers~\cite{Mohan:2014}.

Third generation filters are in most part extensions of the existing second-generation implemented to take advantage of non-local data. The basic idea is to use similar patches in order to estimate and remove high-frequency noise.  Examples of bilateral filter extensions are the NLM~\cite{Buades:2005}, BM3D~\cite{Dabov:2009}. Third generation filters have also been applied to medical imaging in the context of MRI and CT~\cite{Diwakar:2018}.

In the literature, there is not much effort in developing a non-local ADF filter for MRI and CT images~\cite{Mohan:2014, Diwakar:2018}. Only two third generation of ADF were proposed in~\cite{Yang:2013,Yuan:2015}. These extensions are just a trivial addition of non-local patches to the algorithm with no mathematical or empirical selection of optimum parameters.

In~\cite{Yang:2013}, the authors use the original edge-stopping functions. In~\cite{Yuan:2015}, the Tukey's biweight robust estimator was used. Either of them estimated similar patches based on pre-filtered images. The second implementation computed adjacent patches at each iteration of the filter which is unnecessary and computationally expensive.

Therefore, there is a lack of studies of a non-local ADF which is the main subject of this paper.

\section{Anisotropic Diffusion Filter}
\label{sec:anisotropic}

The ADF was proposed by Perona and Malik~\cite{Perona:1990}. It consists in applying concepts of fluid thermodynamics to filter images. Just as molecules with different temperature from its adjacency tends to exchange heat until achieving homogeneous temperature, there would have a propensity to match high-frequency noisy pixel intensities with their local neighbours. In this scenario, image edges are similar to tow large adjacent volumes of fluid with higher capacitance, demanding more time to homogenize the temperature in comparison to the noisy smaller volumes. Therefore, ADF is capable of eliminating the high-frequency noise while preserving image strong edges.

The discrete implementation of ADF is given by Equation~\ref{eq:adf}:
\begin{equation}
I_s^{t+1} \approx I_s^t + \frac{\lambda}{|\eta_s|}\sum_{p \in \eta_s}{g(\nabla I_{s,p}^t,\gamma^t)\nabla I_{s,p}^t}
\label{eq:adf}
\end{equation}
$I^t_s$ stands for the intensity of pixel s
in image $I$ at the instant $t$, $\lambda$ is a diffusion rate scalar, $\gamma^t$ is a positive variable related to the diffusion or smoothing strength which decreases at each iteration, $\eta_s$ represents the set of pixels adjacent to $s$, $g(\cdot)$ is an edge stopping function (ESF) which controls the diffusion process, and $\nabla I^t_{s,p}$ is the image directional gradient from $s$ to $p$ at instant $t$. The directional gradient $\nabla I^t_{s,p}$ can be approximated by $I^t_p - I^t_s$.

The diffusion process follows the ESF which has been widely studied and tested~\cite{Black:1998, Kamalaveni:2015}. Tukey's biweight robust estimator ESF is given in Equation~\ref{eq:tukey} and is an excellent option since it does not affect edges with intensity variation over a certain threshold~\cite{Palma:2014}.
\begin{equation}
g(\nabla I_{s,p}^t,\gamma^t)=\left\{ \begin{array}{l}
\left[1-\left(\frac{\nabla I_{s,p}^t}{\sqrt{5}\gamma^t}\right)^2\right]^2, \nabla I_{s,p}^t \leq \gamma^t \sqrt{5} \\
0, \text{otherwise.}
\end{array} \color{white}\right\}
\label{eq:tukey}
\end{equation}

With respect to parameter $\lambda$, the authors in~\cite{Gerig:1992} proposed the maximum values in order to keep a monotonic variation of the intensities at each iteration, so that no artifacts are inserted to the image. In~\cite{Palma:2014}, the authors corrected one of the constants for 26-adjacency. Table~\ref{tab:lambda} shows the expected maximum values of $\lambda$ and of the fraction $\lambda/|\eta_s|$ according to the number of dimensions of the image ($D$) and to the adjacency size ($|\eta_s|$).

\begin{table}[htbp]
\caption{Maximum ADF values for constant $\lambda$ and expression $\lambda/|\eta_s|$, given the input image dimensions $D$ and the adjacency size $|\eta_s|$.}
\begin{center}
\begin{tabular}{|cc|cc|}
\hline
\boldmath{$D$} & \boldmath{$|\eta_s|$} & \boldmath{$\lambda$} & \boldmath$\lambda/|\eta_s|$ \\
\hline
2 & 4 & 4/5 & 1/5 \\
2 & 8 & 8/7 & 1/7 \\
3 & 6 & 6/7 & 1/7 \\
3 & 18 & 18/13 & 1/13 \\
3 & 26 & 78/47 & 3/47 \\
\hline
\end{tabular}
\label{tab:lambda}
\end{center}
\end{table}

There were several different solutions proposed to estimate the initial $\gamma^0$ for the first iteration and the number of iterations $T$. Some approaches used global and/or local gradient to compute $\gamma$~\cite{Perona:1990, Black:1998, Voci:2004} while others use a planar region (i.e. without strong edges)~\cite{Tsiotsios:2013}.

The approach in~\cite{Palma:2014} used an adaptive method based on both the global gradient and a planar region in order to estimate the best initial $\gamma^0$. While the gradient indicates an upper bound $\gamma_\cal{E}$ to avoid filtering edges, the planar region provides a lower bound $\gamma_\cal{F}$ so that practically all noise is removed. In case of low signal to noise ratio (SNR)\footnote{In the experiments, Gaussian or Rician noise of around 3\% of amplitude is the maximum SNR which is perfectly filtered without blurring significant edges.}, it may happen that $\gamma_\cal{F} > \gamma_\cal{E}$. In these cases, a value between $\gamma_\cal{E}$ and $\gamma_\cal{F}$ should be chosen. Lower $\gamma^0$ translates into a conservative ADF which preserves the stronger edges, while higher $\gamma^0$ constitutes into an aggressive filtering, removing all noise.

Also, since its proposal~\cite{Perona:1990} and as corroborated by~\cite{Voci:2004}, it is very important to update $\gamma^t$ at each iteration. As showed in~\cite{Palma:2014}, it is fundamental to reduce $\gamma^t$ at each iteration in order to preserve the borders and it may be very expensive to recompute $\gamma^t$ at each iteration based on the noise in the planar region and on the strong edges. A conservative proposed solution was to set $\gamma^t = 0.25 \gamma^{t-1}$. That is because it is expected that the strongest noise has at most one-fourth of its adjacent pixels with a similar intensity, while for the edges, it is reasonable to expect that at least one-fourth of the pixels have distinct intensity.  Therefore, reducing $\gamma$ by $0.16 \gamma$ was a conservative adopted criteria\footnote{For $\gamma$ reduction values of other ESFs, please refer to~\cite{Palma:2014}.}.

Finally, with respect to the number of iterations $T$, in~\cite{Tsiotsios:2013} the authors verify the intensity of the strongest edges. That solution is not suitable for decreasing $\gamma^t$ with Tukey's biweight robust estimator given in Equation~\ref{eq:tukey}. In~\cite{Palma:2014}, the authors propose to stop as $\gamma^t < \gamma_{\cal E}/7$ which does not depend on the noise intensity itself.

\section{Non-Local Optimal Anisotropic Diffusion Filter}
\label{sec:method}

The non-local ADF (NL-ADF) proposed in~\cite{Yang:2013,Yuan:2015} do not optimize any filter parameter. We propose a novel non-local ADF with enhanced parameter optimization. It has improvements over the previous optimization procedure proposed in~\cite{Palma:2014}. We also study the non-local patches influence in the filtering process.

\subsection{Local Parameter Optimization}

Instead of estimating the initial $\gamma^0$ based on the standard deviation of edgy or noisy regions, in this paper we propose a more robust and straightforward strategy. Noise pixels in flat regions have a distinct intensity value than all their adjacents. An edge pixel, on the other hand, has a similar intensity to pixels in the same side of the edge and distinct intensity from pixels in the opposite side of the edge. Therefore, instead of computing the standard deviation of pixels to estimate noise, we propose to use the sum of the directional gradients $G_s$ as in Equation~\ref{eq:dir.grad.diff}:
\begin{equation}
G_s = \sum_{p \in \eta_s}{I_s - I_p}
\label{eq:dir.grad.diff}
\end{equation}
Note that the direction of $G_s$ is always computed toward the source pixel $s$. This way, edge pixels will have opposite edge pixel intensities neutralized, which noise pixels will display the highest values of $G_s$. We then select standard deviation ($\sigma_{G_s}$) of the 5$\%$ highest $G_s$ pixels as our $\gamma^0$, as the maximum effect of of the ADF with Tukey's biweight robust estimator happens for $\gamma^0 = \sigma_{G_s}$.

A second contribution is to revise the $\gamma^t$ reduction strategy proposed in~\cite{Palma:2014} based on 1/4 of the adjacent pixels as explained in Section~\ref{sec:anisotropic} as it was a very conservative choice. In this paper, we evaluated faster and slower $\gamma^t$ reductions based on values in range $[0.16 \gamma^{t-1}, 0.96 \gamma^{t-1}]$.

With respect to the stopping criteria a simple alternative is to stop filtering the image as $\gamma^t \leq 0.01 I^{M}$ ($I^{M}$ is the maximum intensity of an image $I$), as 1$\%$ of maximum intensity is imperceptible by humans eyes in medical images.

Finally, diffusion of adjacent pixels should also consider the distance between the source and target pixels. Therefore, we propose the weighted ADF (WADF) based on Equation~\ref{eq:wadf}:
\begin{equation}
I_s^{t+1} \approx I_s^t + \frac{1}{|\eta_s|}\sum_{p \in \eta_s}{\frac{g(\nabla I_{s,p}^t,\gamma^t)}{d_{s,t}}\nabla I_{s,p}^t}
\label{eq:wadf}
\end{equation}
where $d(s,t)$ is the distance between pixels $s$ and $t$. Note that, according to the study proposed in~\cite{Gerig:1992, Palma:2014} with the WADF for a monotonic variation of the intensities at each iteration we may set $\lambda = 1$, eliminating this parameter from the original ADF Equation~\ref{eq:adf}.

\subsection{Non-Local Anisotropic Diffusion Function}

An interesting question overlooked in previous publications is the weight or distance of non-local patches. Most of the papers, including the previous NL-ADF~\cite{Yang:2013,Yuan:2015}, position the non-local patches in a discrete way in an additional third or fourth image dimension. That is, the closest patch is one-pixel-wide distant from the patch of interest. In a similar way, the original NLM~\cite{Buades:2005} uses the spatial distance as a weighting factor, but this is an arbitrary solution. BM3D~\cite{Dabov:2009} and BM4D~\cite{Maggioni:2014} also build a structure with an additional discrete dimension.

As the ADF does not depend on the distance of patches itself, we decided to evaluate patch distance and quantity over the filtering result. The $n$ more similar patches will have the same distance $d$ to the filtered pixel $s$ as a third (or fourth) dimension to a 2D (or 3D) image. Note that $d$ may assume any float value. We will evaluate multiple patch distance variation in a future work.

The choice of the distance $d$ also impacts on the computation of constant $\lambda$. Still, considering the WADF, we may set $\lambda = 1$ even for the non-local weighted anisotropic diffusion filter (NL-WADF) in Equation~\ref{eq:nlwadf}:
\begin{equation}
I_s^{t+1} \approx I_s^t + \frac{1}{|H_s|}\sum_{p \in \eta_s}{\frac{g(\nabla I_{s,p}^t,\gamma^t)}{d_{s,t}}\nabla I_{s,p}^t}
\label{eq:nlwadf}
\end{equation}
$H_s$ comprehends all adjacent pixels to $s$ both local and non-local.

In the case of NL-WADF, the number of pixels with similar intensity to noise and edge pixels should stay proportionally the same. Therefore, we propose also to evaluate the $\gamma^t$ decrease in every iteration using the same range of $[0.32 \gamma^{t-1}, 0.96 \gamma^{t-1}]$.

Also, we determine the similar patches based on filtered images by a median filter~\cite{Manjon:2015}. After this, the patches are used by NL-WADF over the original image.

\section{Experiments}
\label{sec:experiments}

\begin{figure}[!htb]
  \begin{center}
    \begin{tabular}{c}
      \includegraphics[width=7cm]{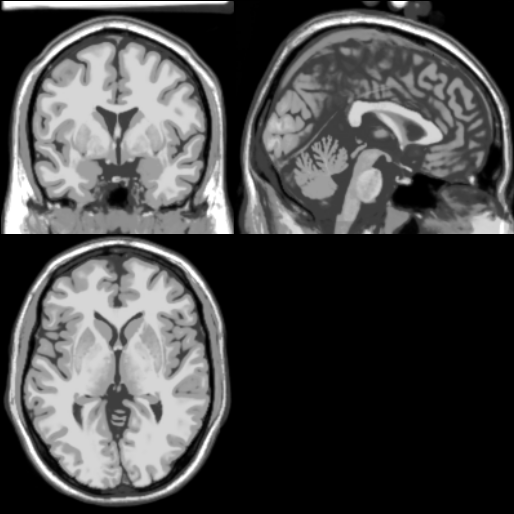} \\
      (a)\\
      \includegraphics[width=7cm]{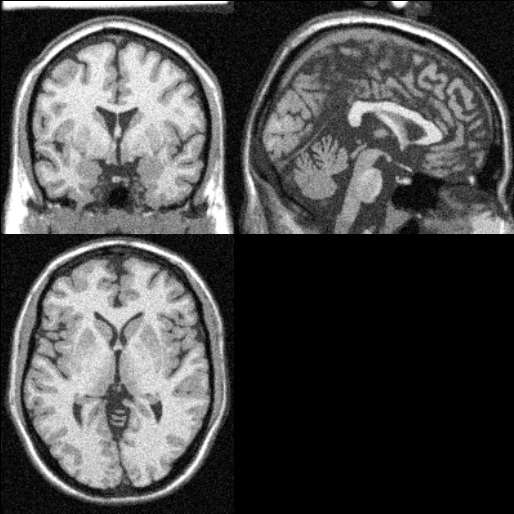} \\
      (b)
    \end{tabular}
  \end{center}
  \caption{Sample slices of BrainWeb Phantom dataset image (a) Ground-truth with no noise; (b) Image corrupted with gaussian noise with 7$\%$ of maximum image intensity.}
  \label{fig:phantom}
\end{figure}

\begin{figure}[!htb]
  \begin{center}
    \begin{tabular}{c}
      \includegraphics[width=6cm]{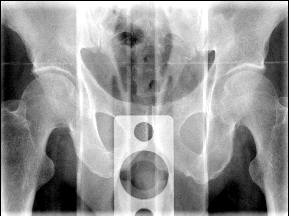} \\
      (a)\\
      \includegraphics[width=6cm]{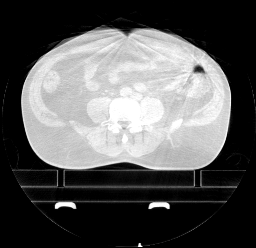} \\
      (b)
    \end{tabular}
  \end{center}
  \caption{Sample slices of radiotherapy images (a) 2D kV planar image; (b) 3D CBCT image.}
  \label{fig:radio}
\end{figure}

The first experiment involves estimating the best parameter values for local and non-local WADF. For this purpose, we used BrainWeb Phantom dataset~\cite{Cocosco:1997} changing the noise intensity between 1 and 9$\%$ of the image maximum intensity. Figure~\ref{fig:phantom} contains slices of the ground-truth image and of an image corrupted by a noise of 7$\%$ of the maximum intensity. The second experiment is a qualitative evaluation in the context of radiotherapy imaging as shown in Figure~\ref{fig:radio}. Images were acquired with Varian TrueBeam and Varian Clinac IX Radiotherapy Systems.

\subsection{Quantitative Evaluation}

We used the following metrics to evaluate the quality of the filtered image in comparison with the ground-truth image with no noise: structural similarity index(SSIM) Equation~\ref{eq:ssim}, mean square error(MSE) Equation~\ref{eq:mse}, peak signal-to-noise ratio(PSNR) Equation~\ref{eq:psnr}, and image quality index(IQI) Equation~\ref{eq:iqi}. We will not present here results for SSIM, since all images achieved accuracy equal to $1.0$. Lower scores are better in terms of MSE and higher scores are better with respect to SSIM and PSNR.

\begin{equation}
SSIM(I,J) = \frac{(2 \mu_I \mu_J + c_1)(2 \sigma_{IJ} + c_2)}{(\mu_I^2+\mu_J^2+ c_1)(\sigma_I^2+\sigma_J^2+ c_2)}
\label{eq:ssim}
\end{equation}
\begin{equation*}
c_1=0.0001 |I||J|
\end{equation*}
\begin{equation*}
c_2=0.0009 |I||J|
\end{equation*}
$I$ and $J$ are the filtered and ground-truth images, respectively, $\mu_I$ and $\sigma_I$ are the mean and standard deviation of pixel intensity in $I$, respectively, and $|I|$ is number of pixels in $I$. $\sigma_IJ$ is the covariance between images $I$ and $J$.
\begin{equation}
MSE(I,J)= \frac{1}{|I|}\sum_s (I_s-J_s)^2
\label{eq:mse}
\end{equation}
\begin{equation}
PSNR(I,J) = 20 * \log( \max(I^{M},J^{M}) ) - 10 \log( MSE(I,J) )
\label{eq:psnr}
\end{equation}
$\max(x,y)$ is the maximum value between x and y. $I^{M}$ is the maximum pixel intensity in $I$.
\begin{equation}
IQI(I,J) = \frac{(4 \mu_I \mu_J \sigma_{IJ})}{(\mu_I^2+\mu_J^2)(\sigma_I^2+\sigma_J^2)}
\label{eq:iqi}
\end{equation}

For the WADF, we tested the $\gamma^t$ reduction rate of $\gamma R = \{0.16\gamma^{t-1}, 0.32\gamma^{t-1}, 0.48\gamma^{t-1}, 0.64\gamma^{t-1}, 0.80\gamma^{t-1}, 0.96\gamma^{t-1}\}$. We also evaluated a conservativeness parameter which sets the initial $\gamma_0 = 0.2\sigma_{G_s}, 0.4\sigma_{G_s}, 0.6\sigma_{G_s}, 0.8\sigma_{G_s}, and \sigma_{G_s}$.

We notice in Table~\ref{tab:local.mri} that the best parameter set for WADF with respect to $\gamma^t$ reduction rate in average to all noise levels is between 0.64$\gamma^{t-1}$ and 0.80$\gamma^{t-1}$ which is much higher than the value adopted in~\cite{Palma:2014}. It is important to mention that the accuracy given by the proposed metrics are not always favorable to 0.64$\gamma^{t-1}$ and 0.80$\gamma^{t-1}$ reduction, but 0.32$\gamma^{t-1}$ provides better results for more conservative filtering(i.e. $\gamma_0 \le 0.4\sigma_{G_s}$).

\begin{table}[!htb]
\caption{Quantitative results of application of WADF over BrainWeb Phantom MRI dataset.}
\centering
\begin{tabular}{|c|ccc|}
   \hline
$\gamma R$  & IQI       & MSE      & PSNR       \\
   \hline
16$\gamma^{t-1}$ & 0.9898 & 3,742 & 27.59 \\
32$\gamma^{t-1}$ & 0.9945 & 2,986 & 28.77 \\
48$\gamma^{t-1}$ & 0.9964 & 2,836 & 29.09 \\
64$\gamma^{t-1}$ & \bf 0.9970 & 2,788 & 29.21 \\
80$\gamma^{t-1}$ & 0.9949 & \bf 2,777 & \bf 29.26 \\
96$\gamma^{t-1}$ & 0.9942 & 2,802 & 29.22 \\
    \hline
\end{tabular}
\label{tab:local.mri}
\end{table}

For the NL-WADF, we tested the following parameters: $\gamma R = \{0.32\gamma^{t-1}, 0.48\gamma^{t-1}, 0.64\gamma^{t-1}, 0.80\gamma^{t-1}, 0.96\gamma^{t-1}\}$; $\gamma_0 = 0.2\sigma_{G_s}, 0.4\sigma_{G_s}, 0.6\sigma_{G_s}, 0.8\sigma_{G_s}, and \sigma_{G_s}$; search radius (SR = 2.0, 3.0 4.0 pixels), non-local patch radius (PR = 1.1, 1.5, 1.8 pixels), the distance from the central pixel of non-local patches (PD = 0.5, 1.0, 2.0 pixels), and the number of non-local patches ($\#$P = 1, 2).

\begin{figure}[!htb]
  \begin{center}
    \begin{tabular}{c}
      \includegraphics[width=7cm]{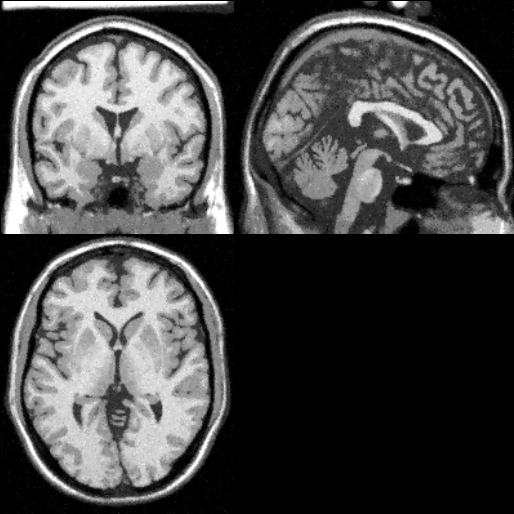} \\
      (a)\\
      \includegraphics[width=7cm]{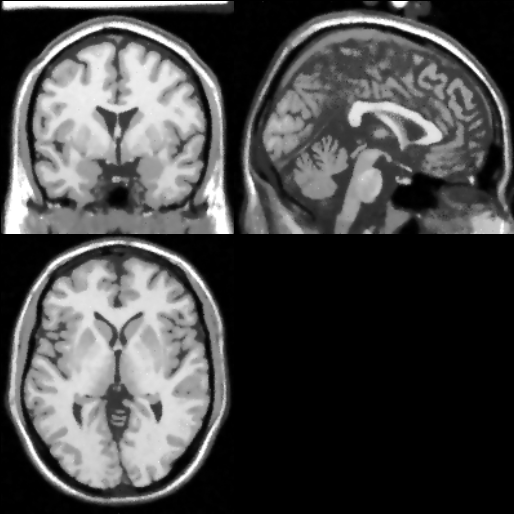} \\
      (b)
    \end{tabular}
  \end{center}
  \caption{Result of filtering image in Figure~\ref{fig:phantom}(b) with (a) conservative NL-WADF and (b) aggressive NL-WADF.}
  \label{fig:phantom.res}
\end{figure}

Figure~\ref{fig:phantom.res} shows sample slices of image showed in Figure~\ref{fig:phantom}(b) filtered conservatively($\gamma^0 = 0.4\sigma_{G_s}$) and aggressively($\gamma^0 = 0.8\sigma_{G_s}$) with parameters SR=3.0, PR=1.9, PD=1.0, $\#$P=2, and $\gamma$R=0.80$\gamma^{t-1}$.

Table~\ref{tab:nonlocal.mri} contains the best results for NL-WADF in terms of the proposed metrics. The best set of parameters in average according to MSE and PSNR is to use SR=4.0, PR=1.1, PD=0.5, $\#$P=2, and a $\gamma$R= 0.32$\gamma^{t-1}$. Note that $\gamma_0$ is not related to the quality of the result, but to the desired conservativeness, which is more related to the goal application. According to IQI, the best set of parameters is SR=3.0, PR=1.9, PD=0.5, $\#$P=1, and $\gamma$R=0.64$\%$.

\begin{table}[!htb]
\caption{Quantitative results of application of NL-WADF over BrainWeb Phantom MRI dataset.}
\centering
\begin{tabular}{|c|ccc|}
\hline
SR; PR; PD; \#R; $\gamma R$    & IQI             & MSE            & PSNR           \\
\hline
2.0; 1.1; 0.5; 2; 0.32 & 0.9968          & 2,719          & 29.46          \\
3.0; 1.1; 0.5; 2; 0.32 & 0.9967          & 2,712          & 29.48          \\
4.0; 1.1; 0.5; 2; 0.32 & 0.9967          & \textbf{2,710} & \textbf{29.49} \\
4.0; 1.9; 1.0; 2; 0.32 & 0.9941          & 3,022          & 28.71          \\
4.0; 1.1; 0.5; 2; 0.48 & 0.9956          & 2,749          & 29.45          \\
2.0; 1.9; 2.0; 1; 0.64 & \textbf{0.9972} & 2,793          & 29.20          \\
3.0; 1.9; 0.5; 1; 0.64 & \textbf{0.9972} & 2,776          & 29.23          \\
4.0; 1.9; 1.0; 2; 0.64 & \textbf{0.9972} & 2,788          & 29.20          \\
     \hline
\end{tabular}
\label{tab:nonlocal.mri}
\end{table}

Note though, that this is by no means the best parameter set for all noise levels and desired conservativeness. Based on the results in Tables~\ref{tab:local.mri} and~\ref{tab:nonlocal.mri} we can also notice that NL-WADF may generate better results as compared to WADF, but this improvement is not statistically significant.

Other important observations are: using one non-local patch produces in 65.7$\%$ of cases a better accuracy than employing two of them and in 69.9$\%$ of the times for more aggressive filtering (i.e. $\gamma_0 \ge 0.6$). In 59.1$\%$ of the times, for a noise level of 7$\%$, using two non-local patches produces a better accuracy. That is the only noise level with better results using 2 non-local patches.

In terms of IQI, $\gamma^t$ reduction of 0.96$\gamma^{t-1}$ produces the higher accuracy in 99$\%$ of the tests, except while dealing with a noise level of 7$\%$.

The analysis of the size of the non-local patch radius is very similar to the number of patches. A radius of 1.1 pixels generates higher accuracy in 52.3$\%$ of times. Smaller radius are a better choice in 62.2$\%$ of times for aggressive filters (i.e $\gamma_0 =0.6\sigma_{G_s}$).

Setting non-local patches at 0.5 pixels from the source patch generates the best results in 55$\%$ of the times and in 61.8$\%$ of the times for more aggressive filtering (i.e. $\gamma_0 \ge 0.6$). This means that non-local patches should have a stronger impact in removing the noise.

With respect to the size of the similar patches search region, a radius of 2.0 pixels is usually sufficient and produces the best results. Conservative parameters are an exception (i.e. $\gamma_0 \le 0.4$) over high-intensity noise in which a radius of 4.0 pixels produces better results. This indicates that a larger search radius is an important parameter to conserve image edges.

\subsection{Qualitative Evaluation}

Using the best parameter set for NL-WADF, we performed a qualitative test with cone beam computed tomography (CBCT) and kV planar radiotherapy images of the low pelvis region provided by hospital AC Camargo in São Paulo/SP, Brazil. A physicist specialist evaluated the quality of the filtered images and their utility for image visualization purposes. 

\begin{figure}[!htb]
  \begin{center}
    \begin{tabular}{c}
      \includegraphics[width=6cm]{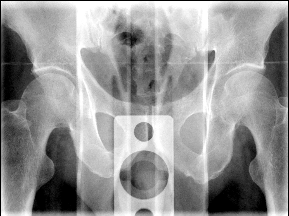} \\
      (a)\\
      \includegraphics[width=6cm]{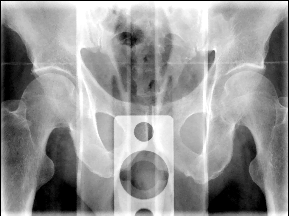} \\
      (b)
    \end{tabular}
  \end{center}
  \caption{Resulting image after filtering image in Figure~\ref{fig:radio}(a) with (a) conservative NL-WADF and (b) aggressive NL-WADF.}
  \label{fig:planar.res}
\end{figure}

\begin{figure}[!htb]
  \begin{center}
    \begin{tabular}{c}
      \includegraphics[width=6cm]{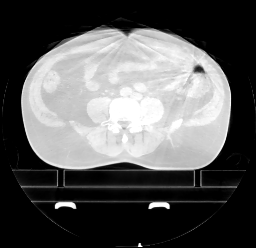} \\
      (a)\\
      \includegraphics[width=6cm]{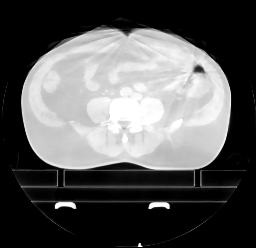} \\
      (b)
    \end{tabular}
  \end{center}
  \caption{Resulting image after filtering image in Figure~\ref{fig:radio}(b) with (a) conservative NL-WADF and (b) aggressive NL-WADF.}
  \label{fig:cbct.res}
\end{figure}

Figures~\ref{fig:planar.res} and~\ref{fig:cbct.res} shows sample slices of filtered planar kV and 3D CBCT images with conservative and aggressive NL-WADF.

Five filtering levels for conservative to aggressive of ten images (3 planar and 7 3D CBCT volumes) were evaluated. The specialist assigned a score from 1 (poor) to 5 (excellent) to each of the filtered images. Table~\ref{tab:radio.result} show the scores. As noticed for radiotherapy visualization purposes, the best filter consists of applying $\gamma^0=0.4\sigma_{G_s}$. More conservative parameters do not remove significantly the noise. Using $\gamma^0>0.4\sigma_{G_s}$ on the other hand on more aggressive filter removes important details in the opinion of the specialist.

\begin{table}[!htb]
\caption{Qualitative evaluation of planar kV and 3D CBCT images by physicist specialist. Grades from 1 (poor) to 5 (excellent) were given to each image. Mean values are presented here.}
\centering
\begin{tabular}{|c|c|}
\hline
$\gamma^0$    & Score   \\
\hline
0.2$\sigma_{G_s}$ & 4.9 \\
0.4$\sigma_{G_s}$ & 5.0 \\
0.6$\sigma_{G_s}$ & 4.9 \\
0.8$\sigma_{G_s}$ & 4.9 \\
$\sigma_{G_s}$    & 4.8 \\
\hline
\end{tabular}
\label{tab:radio.result}
\end{table}

\section{Conclusions}
\label{sec:conclusion}

In this paper, we propose an operational non-local anisotropic diffusion filter for additive noises present in medical images such as MRI and CT. The main contributions of the filter were the distance factor never considered before for non-local patches filtering for ADFs and the weighted contribution of adjacent pixels. Also, we present an extensive evaluation of the parameter values in order to define the best parameter set in terms of SSIM, PSNR, MSE, and DE. Results show that the new non-local version of the ADF is significantly better than the local filter. Finally, we also present a qualitative evaluation of the filter in the context of visual quality of radiotherapy imaging which showed that filtering the image is an important procedure, but the procedure must be conservative avoiding removing important anatomical details.

Future works involve proposing more sophisticated strategies for setting the distance between non-local adjacent patches depending on the source pixel neighbourhood and also based on the neighbourhood similarity.

\section*{Acknowledgment}

We would like to thank AC Camargo and Dr Luiz Juliano Neto for the support with the radiotherapy images. We also thank the training provided by Varian Medical Systems in agreement with the Brazilian Ministry of Health.

\bibliographystyle{IEEEtran}
\bibliography{refs}

\begin{thebibliography}{10}
\providecommand{\url}[1]{#1}
\csname url@samestyle\endcsname
\providecommand{\newblock}{\relax}
\providecommand{\bibinfo}[2]{#2}
\providecommand{\BIBentrySTDinterwordspacing}{\spaceskip=0pt\relax}
\providecommand{\BIBentryALTinterwordstretchfactor}{4}
\providecommand{\BIBentryALTinterwordspacing}{\spaceskip=\fontdimen2\font plus
\BIBentryALTinterwordstretchfactor\fontdimen3\font minus
  \fontdimen4\font\relax}
\providecommand{\BIBforeignlanguage}[2]{{%
\expandafter\ifx\csname l@#1\endcsname\relax
\typeout{** WARNING: IEEEtran.bst: No hyphenation pattern has been}%
\typeout{** loaded for the language `#1'. Using the pattern for}%
\typeout{** the default language instead.}%
\else
\language=\csname l@#1\endcsname
\fi
#2}}
\providecommand{\BIBdecl}{\relax}
\BIBdecl

\bibitem{Beutel:2000}
J.~Beutel, H.~L. Kundel, and R.~L. Van~Metter, \emph{Handbook of medical
  imaging}.\hskip 1em plus 0.5em minus 0.4em\relax Spie Press, 2000, vol.~1.

\bibitem{Mohan:2014}
J.~Mohan, V.~Krishnaveni, and Y.~Guo, ``A survey on the magnetic resonance
  image denoising methods,'' \emph{Biomedical signal processing and control},
  vol.~9, pp. 56--69, 2014.

\bibitem{Diwakar:2018}
M.~Diwakar and M.~Kumar, ``A review on ct image noise and its denoising,''
  \emph{Biomedical Signal Processing and Control}, vol.~42, pp. 73--88, 2018.

\bibitem{Maggioni:2013}
M.~Maggioni, V.~Katkovnik, K.~Egiazarian, and A.~Foi, ``Nonlocal
  transform-domain filter for volumetric data denoising and reconstruction,''
  \emph{IEEE transactions on image processing}, vol.~22, no.~1, pp. 119--133,
  2013.

\bibitem{Rafsanjani:2017}
H.~K. Rafsanjani, M.~H. Sedaaghi, and S.~Saryazdi, ``An adaptive diffusion
  coefficient selection for image denoising,'' \emph{Digital Signal
  Processing}, vol.~64, pp. 71--82, 2017.

\bibitem{Lindenbaum:1994}
M.~Lindenbaum, M.~Fischer, and A.~Bruckstein, ``On gabor's contribution to
  image enhancement,'' \emph{Pattern Recognition}, vol.~27, no.~1, pp. 1--8,
  1994.

\bibitem{Petrou:2010}
M.~Petrou and C.~Petrou, \emph{Image processing: the fundamentals}.\hskip 1em
  plus 0.5em minus 0.4em\relax John Wiley \& Sons, 2010.

\bibitem{Alvarez:1992}
L.~Alvarez, P.-L. Lions, and J.-M. Morel, ``Image selective smoothing and edge
  detection by nonlinear diffusion. ii,'' \emph{SIAM Journal on numerical
  analysis}, vol.~29, no.~3, pp. 845--866, 1992.

\bibitem{Smith:1997}
S.~M. Smith and J.~M. Brady, ``Susan—a new approach to low level image
  processing,'' \emph{International journal of computer vision}, vol.~23,
  no.~1, pp. 45--78, 1997.

\bibitem{Fedorov:2012}
A.~Fedorov, R.~Beichel, J.~Kalpathy-Cramer, J.~Finet, J.-C. Fillion-Robin,
  S.~Pujol, C.~Bauer, D.~Jennings, F.~Fennessy, M.~Sonka \emph{et~al.}, ``3d
  slicer as an image computing platform for the quantitative imaging network,''
  \emph{Magnetic resonance imaging}, vol.~30, no.~9, pp. 1323--1341, 2012.

\bibitem{Buades:2005}
A.~Buades, B.~Coll, and J.-M. Morel, ``A non-local algorithm for image
  denoising,'' in \emph{Computer Vision and Pattern Recognition, 2005. CVPR
  2005. IEEE Computer Society Conference on}, vol.~2.\hskip 1em plus 0.5em
  minus 0.4em\relax IEEE, 2005, pp. 60--65.

\bibitem{Perona:1990}
P.~Perona and J.~Malik, ``Scale-space and edge detection using anisotropic
  diffusion,'' \emph{IEEE Transactions on pattern analysis and machine
  intelligence}, vol.~12, no.~7, pp. 629--639, 1990.

\bibitem{Palma:2014}
C.~A. Palma, F.~A. Cappabianco, J.~S. Ide, and P.~A. Miranda, ``Anisotropic
  diffusion filtering operation and limitations-magnetic resonance imaging
  evaluation,'' \emph{IFAC Proceedings Volumes}, vol.~47, no.~3, pp.
  3887--3892, 2014.

\bibitem{Anand:2010}
C.~S. Anand and J.~S. Sahambi, ``Wavelet domain non-linear filtering for mri
  denoising,'' \emph{Magnetic Resonance Imaging}, vol.~28, no.~6, pp. 842--861,
  2010.

\bibitem{Manjon:2012}
J.~V. Manj{\'o}n, P.~Coup{\'e}, A.~Buades, D.~L. Collins, and M.~Robles, ``New
  methods for mri denoising based on sparseness and self-similarity,''
  \emph{Medical image analysis}, vol.~16, no.~1, pp. 18--27, 2012.

\bibitem{Wong:2004}
W.~C. Wong, A.~C. Chung, and S.~C. Yu, ``Trilateral filtering for biomedical
  images,'' in \emph{Biomedical Imaging: Nano to Macro, 2004. IEEE
  International Symposium on}.\hskip 1em plus 0.5em minus 0.4em\relax Citeseer,
  2004, pp. 820--823.

\bibitem{Xie:2008}
J.~Xie, P.-A. Heng, and M.~Shah, ``Image diffusion using saliency bilateral
  filter,'' \emph{IEEE Transactions on Information Technology in Biomedicine},
  vol.~12, no.~6, pp. 768--771, 2008.

\bibitem{Gerig:1992}
G.~Gerig, O.~Kubler, R.~Kikinis, and F.~A. Jolesz, ``Nonlinear anisotropic
  filtering of mri data,'' \emph{IEEE Transactions on medical imaging},
  vol.~11, no.~2, pp. 221--232, 1992.

\bibitem{Black:1998}
M.~J. Black, G.~Sapiro, D.~H. Marimont, and D.~Heeger, ``Robust anisotropic
  diffusion,'' \emph{IEEE Transactions on image processing}, vol.~7, no.~3, pp.
  421--432, 1998.

\bibitem{Dabov:2009}
K.~Dabov, A.~Foi, V.~Katkovnik, and K.~Egiazarian, ``Bm3d image denoising with
  shape-adaptive principal component analysis,'' in \emph{SPARS'09-Signal
  Processing with Adaptive Sparse Structured Representations}, 2009.

\bibitem{Yang:2013}
M.~Yang, J.~Liang, J.~Zhang, H.~Gao, F.~Meng, L.~Xingdong, and S.-J. Song,
  ``Non-local means theory based perona--malik model for image denosing,''
  \emph{Neurocomputing}, vol. 120, pp. 262--267, 2013.

\bibitem{Yuan:2015}
J.~Yuan, ``Improved anisotropic diffusion equation based on new non-local
  information scheme for image denoising,'' \emph{IET Computer Vision}, vol.~9,
  no.~6, pp. 864--870, 2015.

\bibitem{Kamalaveni:2015}
V.~Kamalaveni, R.~A. Rajalakshmi, and K.~Narayanankutty, ``Image denoising
  using variations of perona-malik model with different edge stopping
  functions,'' \emph{Procedia Computer Science}, vol.~58, pp. 673--682, 2015.

\bibitem{Voci:2004}
F.~Voci, S.~Eiho, N.~Sugimoto, and H.~Sekibuchi, ``Estimating the gradient in
  the perona-malik equation,'' \emph{IEEE Signal Processing Magazine}, vol.~21,
  no.~3, pp. 39--65, 2004.

\bibitem{Tsiotsios:2013}
C.~Tsiotsios and M.~Petrou, ``On the choice of the parameters for anisotropic
  diffusion in image processing,'' \emph{Pattern recognition}, vol.~46, no.~5,
  pp. 1369--1381, 2013.

\bibitem{Maggioni:2014}
M.~Maggioni, V.~Katkovnik, K.~Egiazarian, and A.~Foi, ``Nonlocal
  transform-domain filter for volumetric data denoising and reconstruction,''
  \emph{IEEE transactions on image processing}, vol.~22, no.~1, pp. 119--133,
  2013.

\bibitem{Manjon:2015}
J.~V. Manj{\'o}n, P.~Coup{\'e}, and A.~Buades, ``Mri noise estimation and
  denoising using non-local pca,'' \emph{Medical image analysis}, vol.~22,
  no.~1, pp. 35--47, 2015.

\bibitem{Cocosco:1997}
C.~A. Cocosco, V.~Kollokian, R.~K.-S. Kwan, G.~B. Pike, and A.~C. Evans,
  ``Brainweb: Online interface to a 3d mri simulated brain database,'' in
  \emph{NeuroImage}.\hskip 1em plus 0.5em minus 0.4em\relax Citeseer, 1997.

\end{thebibliography}

\balance

\end{document}